\documentclass[fleqn,usenatbib]{mnras}

\usepackage[T1]{fontenc}

\usepackage{graphicx}	
\usepackage{amsmath}	
\usepackage{amssymb}	
\usepackage{multirow}	
\usepackage{booktabs}

\usepackage{newtxtext,newtxmath}

\newcommand{\heyoka}{\texttt{heyoka}}

\title[Event detection in Taylor methods]{Reliable event detection for Taylor methods in astrodynamics}

\author[F. Biscani et al.]{
Francesco Biscani,$^{1}$\thanks{E-mail: bluescarni@gmail.com}
Dario Izzo$^{2}$
\\
$^{1}$Center for Astronomy (ZAH), University of Heidelberg / Max Planck Institute for Astronomy, Heidelberg, Germany\\
$^{2}$Advanced Concepts Team, European Space Agency (ESA), Noordwijk, The Netherlands
}

\date{Accepted XXX. Received YYY; in original form ZZZ}

\pubyear{2021}

\begin{document}
\label{firstpage}
\pagerange{\pageref{firstpage}--\pageref{lastpage}}
\maketitle

\begin{abstract}
We present a novel approach for the detection of events in systems of ordinary differential equations. The new method
combines the unique features of Taylor integrators with state-of-the-art polynomial root finding techniques to yield
a novel algorithm ensuring strong event detection guarantees at a modest computational overhead. Detailed tests and benchmarks
focused on problems in astrodynamics and celestial mechanics (such as collisional N-body systems, spacecraft dynamics around irregular
bodies accounting for eclipses, computation of Poincar\'e{} sections, etc.) show how our approach is superior in both
performance and detection accuracy to strategies commonly employed in modern numerical integration works. The new algorithm is available in our open source Taylor integration package \heyoka{}.

\end{abstract}




\section{Introduction}
During the numerical integration of ordinary differential equations (ODEs), it is often necessary to
detect the occurrence of specific conditions (or \emph{events}) in the state of a system.
While, in some cases, the detection of an event in and of itself may suffice,
in other cases the system may need to react to the event by discontinuously changing
the state and/or the dynamics. Such systems are known as \emph{hybrid} dynamical systems \citep{Branicky2005}.

Hybrid dynamical systems frequently occur in astrodynamics and celestial mechanics. Examples include:
\begin{itemize}
    \item compact N-body systems, such as young planetary systems, where
    collisions between planetesimals and/or planetary embryos
    play a fundamental role in shaping the long-term dynamical evolution of the system \citep{mordasini2009extrasolar};
    \item spacecraft dynamics around a celestial body in presence of discontinuous eclipse transitions in and out of the shadow
    cone \citep{lang2021spacecraft};
    \item systems governed by physical laws formulated as piecewise functions, such as
    the drag force exerted on a spacecraft in low Earth orbit according to the standard atmospheric model
    \citep{benner1973revised}, or the torques causing planet migration in protoplanetary disks \citep{dtorque}.
\end{itemize}
The more general need for event detection capabilities in non-hybrid systems is also ubiquitous.
In the computation of transit timing variations in exoplanetary systems, for instance, it is necessary
to detect when a planet crosses the line of sight \citep{ttv_nbody}. The construction of
Poincar\'e{} sections in theoretical astrodynamical studies is another example
where event detection capabilities are necessary \citep{teschl2012ordinary}.

The most widely-adopted event detection method in modern ODE integration packages
is based on the detection of changes in the sign of
the event function. After a sign change is detected, the event time can accurately be
located via a root-finding procedure.
The sign change approach is employed in the
widely-used \texttt{SciPy} \citep{scipy}, \texttt{MATLAB} \citep{shampine1997matlab} and \texttt{DifferentialEquations.jl} \citep{diffeqjl} packages.
The popular astrodynamical software \texttt{REBOUND} \citep{rein2012rebound}, while not directly supporting
event detection, includes an example of computation of transit timing variations via sign change monitoring.
More recently, \citet{ttv_nbody} also use sign change monitoring (albeit augmented by the availability of the gradient of the event functions)
for the computation of transit timing variations in N-body simulations.

The sign change method is computationally cheap and easy to implement (in fact, it can be implemented
on top of \emph{any} existing numerical integration library). However, it also suffers from
serious shortcomings, as it does not offer any guarantee that all events are detected (e.g., events may
be missed because they trigger an even amount of times within two sign sampling points). Careful
tuning of the integration timestep using problem-dependent knowledge can reduce the probability
of missing or misdetecting events, but this approach may impose unnecessary restrictions on the timestep selection while
still not being able to ensure that misdetections are completely avoided. The sign change method is thus unsuitable in high-precision
applications where it is of paramount importance that events are reliably detected.

Building on results from a previous paper \citep{hey_mnras}, where we showed how Taylor methods
are competitive with state-of-the-art symplectic and non-symplectic integrators in astrodynamical
applications, we propose in this paper a novel approach for event detection in Taylor integrators.
This new approach combines existing ideas on accurate event detection \citep{shampine1991reliable,shampine2000event,grabner2003integrated}
with the unique features of Taylor integrators, yielding a new algorithm that is
both computationally efficient and reliable.
Our method is based on three main ideas:
\begin{itemize}
    \item the numerical integration of the event functions, which
    are formally added to the ODE system and integrated together with the dynamical equations;
    \item the availability of free dense output in Taylor methods, which yields
    polynomial approximations of the event functions as part of the integration algorithm;
    \item the use of modern polynomial real-root isolation algorithms
    for the fast and reliable localisation of \emph{all} roots of the polynomial
    approximations of the event functions within each integration step.
\end{itemize}
We will show how the resulting algorithm provides strong event detection guarantees at a modest computational overhead.

Our paper is structured as follows: section \S\ref{sec:ed_taylor} recalls the salient features of Taylor integrators and shows how they can be exploited in the context of event detection; section \S\ref{sec:rr_isol} gives an overview of polynomial root finding techniques, with a special focus on real-root isolation schemes; 
section \S\ref{sec:ed_heyoka} focuses  on various details of the implementation of the novel event detection scheme in our Taylor integrator \heyoka{}. In particular, we discuss on the difficult issue of discontinuity sticking for which we present a mathematically-rigorous solution;
section \S\ref{sec:tests_ex} presents a variety of tests, examples and benchmarks in the context of astrodynamics and celestial mechanics; 
section \S\ref{sec:caveats} discusses some caveats and limitations of our approach; 
section \S\ref{sec:concl} draws concluding remarks.

The algorithm described in this paper is implemented in our Taylor integrator \heyoka{}, which is
available as an open source project at the repository \url{https://github.com/bluescarni/heyoka}.

\section{Event detection in Taylor methods}
\label{sec:ed_taylor}
We consider initial-value problems for systems of ordinary differential equations (ODE) in the explicit form
\begin{equation}
\boldsymbol{x}'\left( t \right)=\boldsymbol{F}\left(t, \boldsymbol{x}\left( t \right) \right),\label{eq:ode_00}
\end{equation}
where $t$ is the independent variable and the prime symbol denotes differentiation with
respect to $t$. Taylor's method for the integration of ODEs uses the truncated Taylor series
\begin{equation}
\boldsymbol{x}\left( t_1 \right) = \sum_{n=0}^p \boldsymbol{x}^{\left[ n \right]} \left(t_0\right) h^n\label{eq:taylor_poly_00}
\end{equation}
to propagate the state of the system from $t=t_0$ to $t=t_1$. In eq. \eqref{eq:taylor_poly_00}, $h=t_1-t_0$ is the integration
timestep and
\begin{equation}
a^{\left[ n \right]}\left( t \right) = \frac{1}{n!}a^{\left( n \right)}\left( t \right)
\end{equation}
is a shorthand notation for the normalised derivative of order $n$ with respect to $t$. The coefficients
$\boldsymbol{x}^{\left[ n \right]} \left(t_0\right)$ of the Taylor series \eqref{eq:taylor_poly_00} can be computed
explicitly and efficiently via a process of automatic differentiation. The Taylor order $p$ and integration
timestep $h$ are chosen so that the magnitude of the remainder of the Taylor series is smaller than the desired error
tolerance $\varepsilon$ within the integration timestep. We refer the reader to \citet{hey_mnras} and \citet{jorba2005}
for a detailed description of Taylor's method.

In explicit ODE systems, an event is defined by an equation of the form
\begin{equation}
g\left( t, \boldsymbol{x} \left( t \right) \right) = 0,\label{eq:ev_eq00}
\end{equation}
where the event function $g\left( t, \boldsymbol{x} \left( t \right) \right)$ is sometimes also referred to as the \emph{discontinuity function}.
The goal of event detection is the computation of the solutions of eq. \eqref{eq:ev_eq00} within an integration
timestep, which we refer to as \emph{trigger times}. 
The most widely-adopted event detection method is based on the detection of changes in the sign of
$g\left( t, \boldsymbol{x} \left( t \right) \right)$ (either at the boundaries of a timestep, or between interpolation
points within a timestep if dense output is available). If a sign change is detected,
the trigger time can be computed via a standard root-finding procedure
which involves the repeated propagation of the state of the system 
until eq. \eqref{eq:ev_eq00} is satisfied to the desired accuracy.

The sign change method has a couple of key advantages: it is easy to implement and computationally efficient, especially if events do not
trigger often. On the other hand, it also suffers from serious drawbacks:
\begin{itemize}
    \item it cannot detect events that trigger an even amount of times within
    two sign sampling points (and thus such events will be missed altogether);
    \item the conventional root-finding procedures which are usually coupled to the sign change method
    cannot discern if an event triggered one or multiple times, and may
    thus not converge to the first trigger time in chronological order.
\end{itemize}

In practical terms these shortcomings are not necessarily insurmountable, but they often result in the need to employ application-specific workarounds
to minimise the risk of missing (or misdetecting) events. For instance, if collision detection in planetary N-body systems is cast as an event detection
problem to be solved with the sign change method, then the integration timestep needs to be accurately tuned in order
to minimise the risk of planets entering and exiting the collision within
the same timestep. This situation would lead to the same event triggering twice within one timestep, and thus the collision would be missed altogether by the sign change method. On the other hand, reducing the time step to avoid such a risk, reduces the benefit of an adaptive stepper that would no longer be free to choose longer time steps when possible.

More sophisticated and robust event detection approaches have been proposed in the literature, even though they are not widely used
in modern ODE integration packages
(e.g., see \citet{shampine1991reliable,shampine2000event,grabner2003integrated}). Although the specifics vary, the basic idea is to augment
the original ODE system \eqref{eq:ode_00} with additional differential equations encoding the evolution in time of the left-hand sides of the event equations
\eqref{eq:ev_eq00}:
\begin{equation}
\begin{cases}
\boldsymbol{x}'\left( t \right)=\boldsymbol{F}\left(t, \boldsymbol{x}\left( t \right) \right)\\
\boldsymbol{g}'\left( t \right)=\frac{\partial \boldsymbol{g}}{\partial t}\left( t, \boldsymbol{x} \left( t \right) \right) +
  \frac{\partial \boldsymbol{g}}{\partial \boldsymbol{x}} \left( t, \boldsymbol{x} \left( t \right) \right)
  \cdot \boldsymbol{F}\left(t, \boldsymbol{x}\left( t \right) \right)
\end{cases}.\label{eq:ode_aug00}
\end{equation}
The augmented ODE system \eqref{eq:ode_aug00} is then solved with an integration method providing dense output via polynomial interpolation. Thus, at every
timestep a time polynomial approximation of $\boldsymbol{g}\left( t\right)$ is available, whose zeroes can be determined via root-finding techniques
to yield the trigger times.

This approach is clearly more computationally-intensive than the sign change method, as it involves the solution of additional
differential equations and the computation of the dense output.
However, coupled to a robust polynomial root finding algorithm, the augmented ODE method
can, in principle, guarantee that no events are missed and that multiple occurrences of an event within a timestep are identified in the correct chronological order.

The approach we propose in this paper is similar in spirit to the augmented ODE method, but it takes advantage of the unique features of Taylor's method and of
state-of-the-art polynomial root finding techniques which, to the best of our knowledge, have never been applied before to the problem of event detection.

To begin with, we note how in Taylor methods it is not necessary to compute the explicit analytical expression of $\boldsymbol{g}'\left( t \right)$
in the formulation of the augmented ODE system \eqref{eq:ode_aug00}. Indeed, at each timestep we can build the Taylor series expansion
\begin{equation}
\boldsymbol{g}\left( \tau \right) = \sum_{i=0}^p \boldsymbol{g}^{\left[ i \right]}\left( t_0 \right) \tau^i
\label{eq:ev_taylor_01}
\end{equation}
(where $\tau \in \left[ 0, h\right]$) \emph{directly}
from the derivatives
of the state variables $\boldsymbol{x}^{\left[ n \right]} \left(t_0\right)$ (which are available at no added cost as in Taylor's method
we are solving the dynamical equations \eqref{eq:ode_00}) via a process of automatic differentiation.
We refer to \citet[\S 2.1]{hey_mnras} and \citet[\S 2.2]{jorba2005} for detailed examples of automatic differentiation
in the context of Taylor methods.

Secondly, we recall that the Taylor polynomial \eqref{eq:ev_taylor_01} is a high-fidelity approximation of the time evolution
of the left-hand side of the event equation \eqref{eq:ev_eq00} within the current timestep. In other words, in Taylor methods, the required dense output
is already built in polynomial form as part of the numerical integration process, i.e.,
it does not require additional computations and it is guaranteed to approximate the solution to the desired tolerance within
the integration timestep. Thus, both the computational
overhead and potential accuracy issues of polynomial interpolations
that characterise previous implementations of the augmented ODE 
approach are completely avoided in Taylor methods.

Lastly, whereas existing implementations of the augmented ODE method (mostly deployed on Runge-Kutta type of solvers) use
Sturm sequences for the computation of polynomial roots,
in our implementation we propose the use of a more modern and computationally-efficient algorithm, based on Descartes' rule of signs.
This algorithm will be described in the next section, where we will also give a brief general overview of polynomial root finding techniques.

\section{Real-root isolation}
\label{sec:rr_isol}
Polynomial root finding is a long-standing problem that has been the subject of much research throughout history. The Abel-Ruffini
theorem states that there is no solution in radicals to general polynomial equations of degree higher than four.
Because Taylor methods are often used in high-precision setups, the order $p$ in eq. \eqref{eq:ev_taylor_01} is typically
much greater than four, and thus in practice, for event detection purposes, we must resort to numerical root-finding approaches.

Numerical algorithms for polynomial root finding can be broadly divided into three categories:
\begin{itemize}
    \item methods which locate \emph{all} polynomial roots at the same time using complex arithmetic,
    such as the Durand-Kerner \citep{kerner} and Aberth \citep{aberth} methods;
    \item methods which locate one root at a time, such as Newton's method or other general-purpose iterative
    algorithms;
    \item methods which locate roots in a specific region of the complex plane or the real line.
\end{itemize}

The methods which locate all roots at the same time are numerically robust and easy to implement. However, because they require the use of
complex arithmetic, it may be difficult to decide whether a root with a small imaginary part is real or not. Moreover, computing all roots
when we are interested only in real roots in a specific interval will result in a waste of computational resources.

Derivative-based iterative methods, such as Newton's, perform very well, but they are highly sensitive to the choice of initial guess
and in general they cannot determine how many real roots exist in a specific interval on the real line.

The third category of methods focuses on the computation of disjoint intervals on the real line, called isolating intervals, each containing 
exactly one real root. This computation is called \emph{real-root isolation}. After determining isolating intervals, one can use fast numerical
methods, such as
Newton's, for improving the precision of the result.

Real-root isolation algorithms are an ideal fit for event detection purposes because they provide real-valued results (while
using only real-valued arithmetic) and they are guaranteed to locate all the real zeroes of a polynomial within an interval.
The oldest real-root isolation algorithm uses Sturm sequences \citep{sturm}, but
modern algorithms based on Descartes' rule of sign are more computationally-efficient \citep{kobel2016computing, sagraloff2016computing}.
In the next section, we will give
a brief overview of the fastest known real-root isolation algorithm, due to \citet{ca76}.

\subsection{The Collins-Akritas algorithm}
Descartes' rule of signs states that the number $N_{+}$ of \emph{positive real roots} in a polynomial (counted with their multiplicity)
is at most the number of sign changes
$N_\textnormal{sc}$ in the sequence of coefficients (omitting zero coefficients), and that the difference
$N_\textnormal{sc} - N_{+}$ is a nonnegative even integer.
This implies, in particular, that if $N_\textnormal{sc}$ is zero or one, then there are
exactly zero or one positive roots, respectively. For instance, in the polynomial
\begin{equation}
x^3+x^2-x-1,
\end{equation}
the sequence of coefficient signs is $\left( +,+,-,-\right)$, i.e., the coefficients change sign only once. Thus, $N_\textnormal{sc}=1$
and the polynomial has exactly one positive real root.

Descartes' rule of signs can be extended to provide information about real zeroes in an interval other than $\left( 0, +\infty \right)$
via polynomial transformations (e.g., Budan's theorem uses linear changes of variable to provide a result similar to Descartes' rule of sign
valid for an arbitrary interval \citep{budan}).
Despite being foundational results in the theory of polynomial equations, Descartes' rule of signs and its generalisations are
typically not immediately useful
in practice, because if $N_\textnormal{sc} > 1$ (as it is often the case) then the exact number of positive 
real roots cannot be determined.

The main idea behind the Collins-Akritas (CA) algorithm \citep{ca76} is that of iteratively splitting an interval
in which we are interested to locate the real roots of a polynomial into smaller subintervals, and to recursively apply Descartes' rule of
sign on the subintervals until $N_\textnormal{sc}$ is 0 or 1 for all subintervals. The subintervals for which $N_\textnormal{sc} = 1$
are then the isolating intervals. In order to apply Descartes' rule of signs on
a subinterval, the original polynomial must be transformed so that the subinterval range is mapped
to the $\left( 0, +\infty \right)$ range
in the 
transformed polynomial. In order to achieve this, at each iteration the CA algorithm applies the following polynomial changes of variables:
\begin{itemize}
    \item scaling by two: $x \rightarrow \frac{x}{2}$;
    \item translation by 1: $x \rightarrow x + 1$;
    \item inversion: $x \rightarrow \frac{1}{x}$.
\end{itemize}
The first change of variables (scaling by two) is used to bisect an interval into two smaller subintervals. The latter two changes
of variables constitute a homographic (or M\"{o}bius) transformation used to map a (finite) subinterval range to
the $\left( 0, +\infty \right)$ range. The key insight of the CA algorithm is that, thanks to a theorem by Vincent,
a finite sequence
of bisections and homographic transformations is guaranteed to eventually result in a polynomial for which $N_\textnormal{sc}$ is 
0 or 1, thus ensuring the termination of the algorithm with the identification of all isolating intervals in a finite number of steps
\citep{vincent,ostrowski1950note,obreschkoff1963verteilung}.

\subsection{Computational complexity}
\label{subsec:complexity}
From a computational complexity point of view, the most expensive step of the CA algorithm is the translation $x \rightarrow x + 1$
(which, combined with the inversion $x \rightarrow \frac{1}{x}$, results in a homographic transformation).
Whereas the scaling by two and the inversion have linear complexity $\operatorname{O}\left( p \right)$ in the polynomial degree $p$, the
translation step has quadratic complexity $\operatorname{O}\left( p^2 \right)$ (as can be seen by a simple application of the
binomial theorem).
In the application of the CA
algorithm to event detection, it is always necessary to perform at least one initial homographic
transformation to map the original time interval $\left[ 0, h \right]$
to the $\left( 0, +\infty \right)$ range. Thus, at the very minimum, the application of the CA algorithm
results in a step of computational complexity $\operatorname{O}\left( p^2 \right)$ in the Taylor order $p$.

If the application of Descartes' rule of signs to the transformed polynomial immediately yields $N_{sc} = 0$, then
no event takes place in the time interval $\left[ 0, h \right]$ and no further computations are necessary.
On the other hand, as shown in \citet{jorba2005}, the process of automatic differentiation necessary
for an efficient implementation of Taylor
methods also has quadratic complexity $\operatorname{O}\left( p^2 \right)$ in the Taylor order $p$.
Thus, when $N_{sc} = 0$ after the initial transformation, event detection via the CA algorithm does not worsen the computational complexity
of Taylor methods and adds only a modest overhead that scales linearly with the number of events to be detected.

If $N_{sc} = 1$, then the interval $\left[ 0, h \right]$ is guaranteed to contain exactly one real root, which can be located
via a standard iterative root-finding method (e.g., Newton's). Because $\left[ 0, h \right]$ already provides a bounding interval
for the root, locating the root at high precision typically necessitates only a handful of iterations, each requiring
the evaluation of a polynomial and, depending on the algorithm, its derivative.

If $N_{sc} > 1$, then the interval $\left[ 0, h \right]$ is bisected into two halves. Each half is subject to homographic transformations
that generate two polynomials, and the same steps described above are then applied recursively to the two new polynomials.

It is thus clear that, in order to be optimally efficient for event detection purposes, the CA algorithm should:
\begin{itemize}
    \item immediately return $N_{sc} = 0$ after the initial homographic transformation if no event triggers within
    a timestep,
    \item immediately return $N_{sc} = 1$ after the initial homographic transformation if an event triggers once within
    a timestep,
    \item require the minimum amount of bisections if an event triggers multiple times within a timestep.
\end{itemize}
Although these optimality conditions are not implied by the CA algorithm (which only guarantees termination after a finite number
of steps), in \S\ref{sec:tests_ex} we will show how, in our tests, for event detection purposes in Taylor methods the CA algorithm
behaves almost always in the optimal way.

\subsection{Fast exclusion testing}
\label{subsec:fast_excl}
If we assume that events are relatively rare (i.e., in most integration timesteps no event triggers), then it is possible to use
fast exclusion tests to improve performance with respect to the quadratic complexity of the CA algorithm. The idea is to compute
an \emph{enclosure} for the event polynomial \eqref{eq:ev_taylor_01}, that is, an interval within which the values of
the polynomial are guaranteed to lie for the duration of the timestep. If the enclosure does not include zero, then the event polynomial
cannot have real roots within the integration timestep. Because the complexity for the computation of the enclosure is linear in the polynomial
order, a fast exclusion test performed before the application of the CA algorithm can lead to noticeable speedups (especially
for extended-precision integrations, where the polynomial order is very high).

\citet[\S 4.2.2]{park1996state} propose a fast polynomial enclosure computation algorithm which uses only floating-point additions
(see also \citet{neumaier1990interval}). Our experiments
however indicate that this approach is not effective for high-order Taylor integrators, because it leads to a high rate of false positives (i.e., the
inclusion test indicates the possible existence of a root which is not actually present). Thus, in \heyoka{}
we implement the more straightforward approach of evaluating the event polynomial \eqref{eq:ev_taylor_01} via interval arithmetic, using the
interval $\left[ 0, h \right]$ as evaluation point. This approach, while more computationally intensive than the enclosure computation proposed by
\citet{park1996state}, is able to reliably exclude the occurrence of events (as we will show in \S\ref{sec:tests_ex}).

\section{Event detection in \texttt{\sc heyoka}}
\label{sec:ed_heyoka}
In this section we will focus on the explanation of several important aspects of the practical
implementation of event detection in our open source Taylor integrator \heyoka{}.

\subsection{Timestep selection}
\label{subsec:t_select}
As explained in detail in \citet[\S 2.2]{hey_mnras}, \heyoka{} uses an estimation of the size of the remainders of the Taylor series \eqref{eq:taylor_poly_00}
of the dynamical equations to quantify the error committed at each step of the integration.
The timestep size is then selected so that the magnitude of the remainders
is less than the desired error tolerance $\varepsilon$ (either in an absolute or relative sense).

When event detection is activated, the Taylor series \eqref{eq:ev_taylor_01} of the event functions are considered together with the Taylor series of the
dynamical equations \eqref{eq:taylor_poly_00} in the determination of the timestep size.
In other words, \heyoka{} ensures that the time evolution of both the dynamical variables and the event functions is solved with the same error tolerance $\varepsilon$,
so that the accuracy of event detection is consistent with the overall accuracy of the integrator.

The inclusion of the event functions in the determination of the timestep size is necessary in order to ensure accurate event detection. However, an important 
consequence must be emphasised: event functions whose integration requires very small timesteps will slow down a numerical integration
that could otherwise be carried out with larger timesteps when considering only the dynamical equations (see the discussion in
\S\ref{subsec:ecl_traj} for a concrete example).

\subsection{Terminal and non-terminal events}
\label{subsec:t_nt_ev}
Event detection in \heyoka{} makes a fundamental distinction between terminal and non-terminal events.

Non-terminal events
do not change the system’s dynamics, nor do they alter the state vector of the system. A typical use of non-terminal events is to detect
and log when the system reaches a particular state of interest (e.g., flagging close encounters between celestial bodies, detecting when a velocity or 
coordinate is zero, etc.). Each non-terminal event is associated to a user-defined callback function that will be invoked when the event triggers.

By contrast, terminal events do modify the dynamics and/or the state of the system, and they can thus be used in the implementation
of hybrid dynamical systems.  A typical example of terminal event is the (in)elastic
collision of two bodies,
which instantaneously and discontinuously changes the bodies' velocity vectors.
Another example is the switching on of a spacecraft engine, which alters the differential equations governing the dynamics of the spacecraft. Like in the
case of non-terminal events, a user-defined callback function to be executed at the trigger time can be associated to a terminal event.

At every integration timestep, \heyoka{} performs event detection for both terminal and non-terminal events in the manner described in
\S\ref{sec:rr_isol}: the isolating intervals for all events are identified via the CA algorithm, and a numerical root finding procedure is employed
to accurately locate the zero within each isolating interval (in \heyoka{} we use the TOMS 478 method from \citet{alefeld1995algorithm},
which is available in the Boost C++ libraries \citep{karlsson2005beyond}). If one or more terminal events are detected, within the timestep,
\heyoka{} will sort the detected terminal events by time and will select the first terminal event triggering in chronological order (or reverse chronological 
order when integrating backwards in time). All the other terminal events and all the non-terminal events triggering \emph{after} the first terminal event are 
discarded. \heyoka{} then propagates the state of the system up to the trigger time of the first terminal event, executes the callbacks of the surviving 
non-terminal events in chronological order and finally executes the callback of the first terminal event.

\subsection{Handling discontinuity sticking}
\label{subsec:d_stick}
A difficult numerical issue arises when the numerical integration stops and then resumes in correspondence of a terminal event.
Due to both floating-point rounding and to the nonzero error tolerance $\varepsilon$ of the integration scheme, when the integration restarts
the \emph{same} terminal event that caused the integrator to stop may re-trigger, forcing the integrator to stop again immediately.
Even worse, this process may continue in an endless loop
that would prevent the integrator from making any progress. In the literature, this phenomenon is known as \emph{discontinuity sticking}
\citep{park1996state, mao2002efficient}. In this subsection, we will first analyse the problem from a mathematical standpoint and we will then explain
the solution we adopted in \heyoka{} to deal with this issue.

For simplicity, we consider an ODE system with a single state variable $x\left( t \right)$ and a single event
equation $g\left( x\left( t \right) \right)$ (the generalisation to the multivariate case is trivial and here not considered
as it only complicates the notation, hiding the underlying point we are trying to make).

Within an integration timestep the infinite Taylor series expansions of the event function and of the state variable read:
\begin{align}
{g}\left( \tau \right) &= \sum_{i=0}^\infty {g}^{\left[ i \right]}\left( t_0 \right) \tau^i,\\
{x}\left( \tau \right) &= \sum_{i=0}^\infty {x}^{\left[ i \right]}\left( t_0 \right) \tau^i,
\end{align}
where $t_0$ is the time at the beginning of the timestep and $\tau$ is the time coordinate
relative to $t_0$. In a Taylor integrator, these expansions are stopped at order $p$:
\begin{align}
{g}\left( \tau \right) &= \sum_{i=0}^p {g}^{\left[ i \right]}\left( t_0 \right) \tau^i + {r}_g\left( \tau \right) =
\tilde{{g}}\left( \tau \right) + {r}_g\left( \tau \right), \\
{x}\left( \tau \right) &= \sum_{i=0}^p {x}^{\left[ i \right]}\left( t_0 \right) \tau^i + {r}_x\left( \tau \right) =
\tilde{{x}}\left( \tau \right) + {r}_x\left( \tau \right),
\end{align}
where the remainders ${r}_g$ and ${r}_x$ are both guaranteed to remain smaller in magnitude than the integration tolerance $\varepsilon$
for the duration of the timestep, i.e., $\forall \tau \le h$:
\begin{equation}
\begin{aligned}
\left| {r}_g\left( \tau \right) \right| & < \varepsilon,\\
\left| {r}_x\left( \tau \right) \right| & < \varepsilon.\\
\end{aligned}
\end{equation}
Event detection is performed on the truncated series $\tilde{{g}}\left( \tau \right)$, yielding a trigger time $\tilde{\tau}_e$ such that
\begin{equation}
\tilde{{g}}\left( \tilde{\tau}_e \right) = 0
\end{equation}
and
\begin{equation}
\left| {g}\left( \tilde{\tau}_e \right) \right| = \left| {r_g}\left( \tilde{\tau}_e \right) \right| < \varepsilon.
\end{equation}
If event detection were performed on the untruncated Taylor expansion instead, it would yield another time $\tau_e$ such that
\begin{equation}
{g}\left( \tau_e \right) = 0.
\end{equation}

\begin{figure}
\begin{center}
 \includegraphics[width=.8\columnwidth]{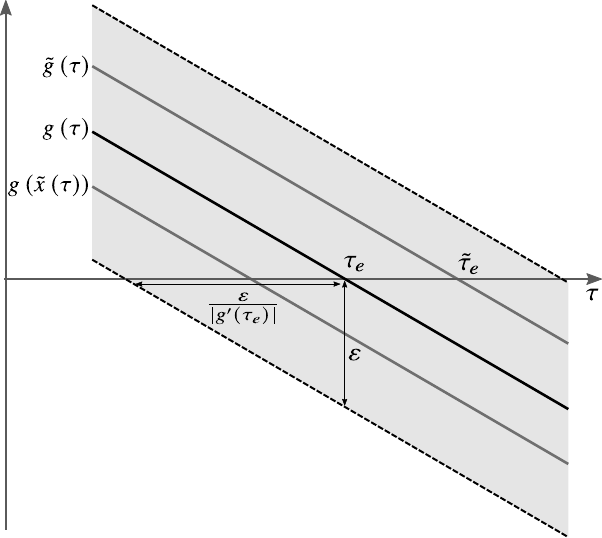}
\end{center}
 \caption{Estimating the uncertainty on the determination of a zero of an event function in the linear approximation.
 The exact event function $g\left( \tau \right)$ has
 a zero for $\tau=\tau_e$. Event detection is however performed on the truncated Taylor series of the event function $\tilde{g}\left( \tau \right)$,
 yielding a value of $\tau=\tilde{\tau}_e \neq \tau_e$. The adaptive timestep deduction scheme ensures that
 $\left| g\left(\tau\right) - \tilde{g}\left(\tau\right) \right| < \varepsilon$ within the integration step, where $\varepsilon$
 is the integrator tolerance. The value of the event function when the integration is restarted after the detection of an event
 is ${g}\left( \tilde{{x}} \left( \tilde{\tau}_e\right)\right)$. Both
 $\tilde{g}\left(\tau\right)$ and ${g}\left( \tilde{{x}} \left( \tau \right)\right)$ do not differ from $g\left(\tau\right)$ more than
 $\varepsilon$ in absolute value, and thus
 $\left| {g}\left( \tilde{{x}} \left( \tilde{\tau}_e\right)\right) \right| < 2\varepsilon$.
  In the linear approximation around the zero of the event function,
 the uncertainty $\varepsilon$ on the vertical axis can be projected on the $\tau$ axis
 via the derivative of the event function, thus yielding an estimation of the time interval within which discontinuity sticking can occur.
 }
 \label{fig:event_zero}
\end{figure}

After the detection of an event, the state of the system has to be propagated
up to the detected event trigger time  $\tilde{\tau}_e$. The problem of discontinuity sticking can emerge at  this stage. Due to both floating-point rounding
and to the nonzero error tolerance $\varepsilon$ of the integration scheme,
the propagated state $\tilde{x}(\tilde \tau_e)$ will not satisfy exactly the event equation \eqref{eq:ev_eq00}.
As a consequence, during the \emph{following} timestep, the newly expanded event function may have a detectable root representing the same event again. This
potentially leads to cases where a never-ending loop of re-detection of the same event (discontinuity sticking) is created.

In mathematical terms, the value of the event function at the beginning of the following
timestep is ${g}\left( \tilde{{x}} \left( \tilde{\tau}_e\right)\right)$ (because we used the truncated Taylor series $\tilde{{x}} \left( \tau\right)$
to propagate the state of the system). Because both $\tilde{g}\left( \tau \right)$ and ${g}\left( \tilde{{x}} \left( \tilde{\tau}\right)\right)$
are approximations of $g\left(\tau\right)$ with an accuracy of order $\varepsilon$ (see the pictorial sketch in Figure \ref{fig:event_zero}),
we can then infer that
\begin{equation}
\left| {g}\left( \tilde{{x}} \left( \tilde{\tau}_e\right)\right) \right| < 2\varepsilon.
\end{equation}

It is now easy to see how, in the linear approximation around the zero of the event function,
we can formulate an estimation of the time interval $\Delta \tau_d$ within which discontinuity sticking
can occur (see Figure \ref{fig:event_zero}):
\begin{equation}
\Delta\tau_d < \frac{2\varepsilon}{\left| {g}^\prime\left( \tau_e \right) \right|}.\label{eq:cd_heur_00}
\end{equation}
Note that the exact value of ${g}^\prime\left( \tau_e \right)$ is not known during the integration, however it is approximated
to accuracy $\sim \varepsilon$ by $\tilde{g}^\prime\left( \tilde{\tau}_e \right)$. We can then quantify the relative error we commit
when estimating $\Delta\tau_d$ via $\tilde{g}^\prime$ rather than ${g}^\prime$:
\begin{equation}
\frac{\frac{2\varepsilon}{{g}^\prime}-\frac{2\varepsilon}{{g}^\prime+\varepsilon}}{\frac{2\varepsilon}{{g}^\prime}}=\frac{\varepsilon}{{g}^\prime+\varepsilon}.
\end{equation}
Thus, in the common case $\varepsilon \ll {g}^\prime$, the relative error is $\varepsilon/{g}^\prime$, while in the extreme case $\varepsilon \gg {g}^\prime$
the relative error on $\Delta\tau_d$ is $\sim 1$. Thus, we can add an extra factor $\times 2$ to the estimate \eqref{eq:cd_heur_00}
as an additional safety margin.

In \heyoka{}, whenever a terminal event triggers, it enters a \emph{cooldown} period of time within which the event is not allowed to trigger again.
By default the cooldown value is computed via eq. \eqref{eq:cd_heur_00}, but the user can override the deduction logic by providing an explicit value
for the cooldown (including zero). In \S\ref{subsec:ecl_traj}, we will show how the cooldown mechanism is effective in dealing with the issue
of discontinuity sticking.

\section{Tests and examples}
\label{sec:tests_ex}

\subsection{The Hénon \& Heiles lost integral}
\label{subsec:third_int}
As a first example, to test and validate \heyoka{}'s event detection machinery, we will reproduce here a famous numerical experiment by \citet{henon1964applicability} investigating the existence
of a third integral of motion in axisymmetric potentials. On the basis of early numerical experiments and theoretical analyses (e.g., see
\citet{contopoulos1963existence}),
it was conjectured that axisymmetric potentials feature an additional integral of motion (beside the energy and the angular momentum integral),
thus implying the Liouville-integrability of all axisymmetric systems (including, e.g., cubic galactic potentials).

In their analysis, \citet{henon1964applicability} consider the autonomous Hamiltonian system
\begin{equation}
\mathcal{H}\left(\dot{x}, \dot{y}, x, y \right) = \frac{1}{2}\left(\dot{x}^2+\dot{y}^2\right) + U\left(x, y\right),\label{eq:as_ham}
\end{equation}
where the potential $U\left(x, y\right)$ is defined as
\begin{equation}
U\left(x, y\right) = \frac{1}{2}\left( x^2 + y^2 + 2x^2y - \frac{2}{3}y^3 \right).
\end{equation}
$x$ and $y$ are the cylindrical coordinates of a point mass moving under the influence of the
axisymmetric potential $U\left(x, y\right)$. The angular momentum integral of motion has already been eliminated in \eqref{eq:as_ham} via the introduction
of cylindrical coordinates, and thus we are left with a two degrees of freedom system featuring one known integral of motion (the Hamiltonian itself).

For a given energy level $\mathcal{H}$, eq. \eqref{eq:as_ham} imposes the following constraint:
\begin{equation}
\frac{1}{2}\dot{x}^2 = \mathcal{H} - U\left(x, y\right) - \frac{1}{2}\dot{y}^2 \geq 0,\label{eq:as_constr_00}
\end{equation}
as $\frac{1}{2}\dot{x}^2$ cannot assume negative values. The inequality \eqref{eq:as_constr_00} defines a \emph{volume} in the 3D phase space
$\left(x, y, \dot{y}\right)$. On the other hand, if another integral of motion $\mathcal{C} = \mathcal{C}\left( \dot{x}, \dot{y}, x, y \right)$ exists,
we can invert it to determine $\dot{x} = \dot{x}\left( x, y, \dot{y}, \mathcal{C} \right)$, plug this definition of $\dot{x}$ into eq. \eqref{eq:as_ham}
and obtain an equation of type
\begin{equation}
f\left(x, y, \dot{y}, \mathcal{H}, \mathcal{C}\right) = 0. \label{eq:2d_surf}
\end{equation}
Eq. \eqref{eq:2d_surf} defines a \emph{surface} embedded in the 3D phase space $\left(x, y, \dot{y}\right)$, rather than a volume. Thus,
\citet{henon1964applicability} propose the use of Poincar\'e{} sections \citep{teschl2012ordinary} to investigate the existence of $\mathcal{C}$:
if $\mathcal{C}$ exists, then the Poincar\'e{} sections will consist of closed curves, otherwise they will resemble densely-filled surfaces.

Following \citet{henon1964applicability}, for the computation of Poincar\'e{} sections we consider the intersections of the solutions
of \eqref{eq:as_ham} with the $\left(y, \dot{y}\right)$ plane in the 3D phase space $\left(x, y, \dot{y}\right)$,
and we consider only intersections for which $\dot{x} > 0$ (i.e., the particle is moving from below to above the plane). The intersection
of a solution with the $\left(y, \dot{y}\right)$ plane can be detected in \heyoka{} via the trivial event equation
\begin{equation}
x = 0.
\end{equation}
When a zero of the event function is detected, the values of $y$ and $\dot{y}$ at the intersection time
are then determined via the dense output provided by the Taylor series of the solution and added to the Poincar\'e{} section.

\begin{figure*}
 \includegraphics[width=\textwidth]{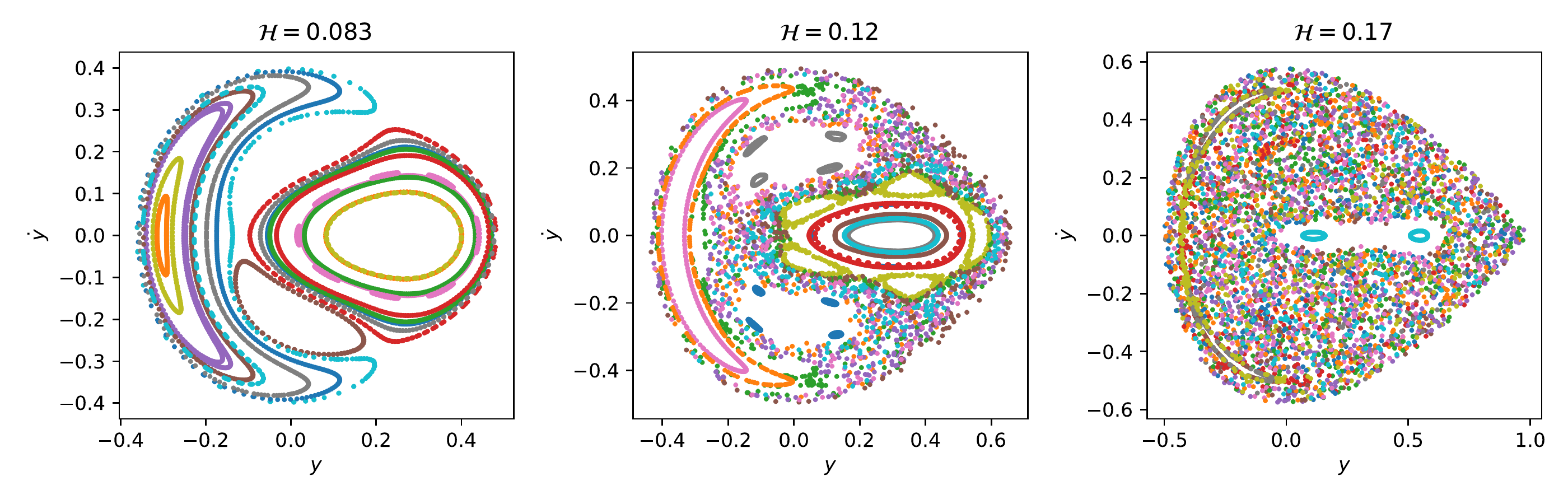}
 \caption{Poincar\'e{} sections on the $\left(y, \dot{y}\right)$ plane for the axisymmetric potential studied
 by \citet{henon1964applicability}
 (see \S \ref{subsec:third_int}). Three energy levels $\mathcal{H}$ are considered. The Poincar\'e{} section for the lowest energy level
 $\mathcal{H} = 0.083$ (left panel) consists of closed curves and would thus seem to indicate the existence of an additional integral of motion
 (beside the energy).
 However, as the energy level increases (center panel), regions of chaotic motion start to appear. Eventually, at the highest energy level
 (right panel) chaotic motion dominates the portrait, thus indicating that the additional integral of motion does not exist.}
 \label{fig:poinc}
\end{figure*}

The Poincar\'e{} sections for the three energy levels considered by \citet{henon1964applicability} are visualised in Fig. \ref{fig:poinc}. Whereas
at low energy levels the Poincar\'e{} section indeed resembles a collection of closed curves, as the energy level increases chaotic regions start to appear,
and eventually dominate the portrait at the highest energy level. These numerical results led \citet{henon1964applicability} to conclude that no additional
integrals of motion exist in general for axisymmetric potentials, thus giving a negative answer to earlier conjectures.

\begin{figure}
 \includegraphics[width=\columnwidth]{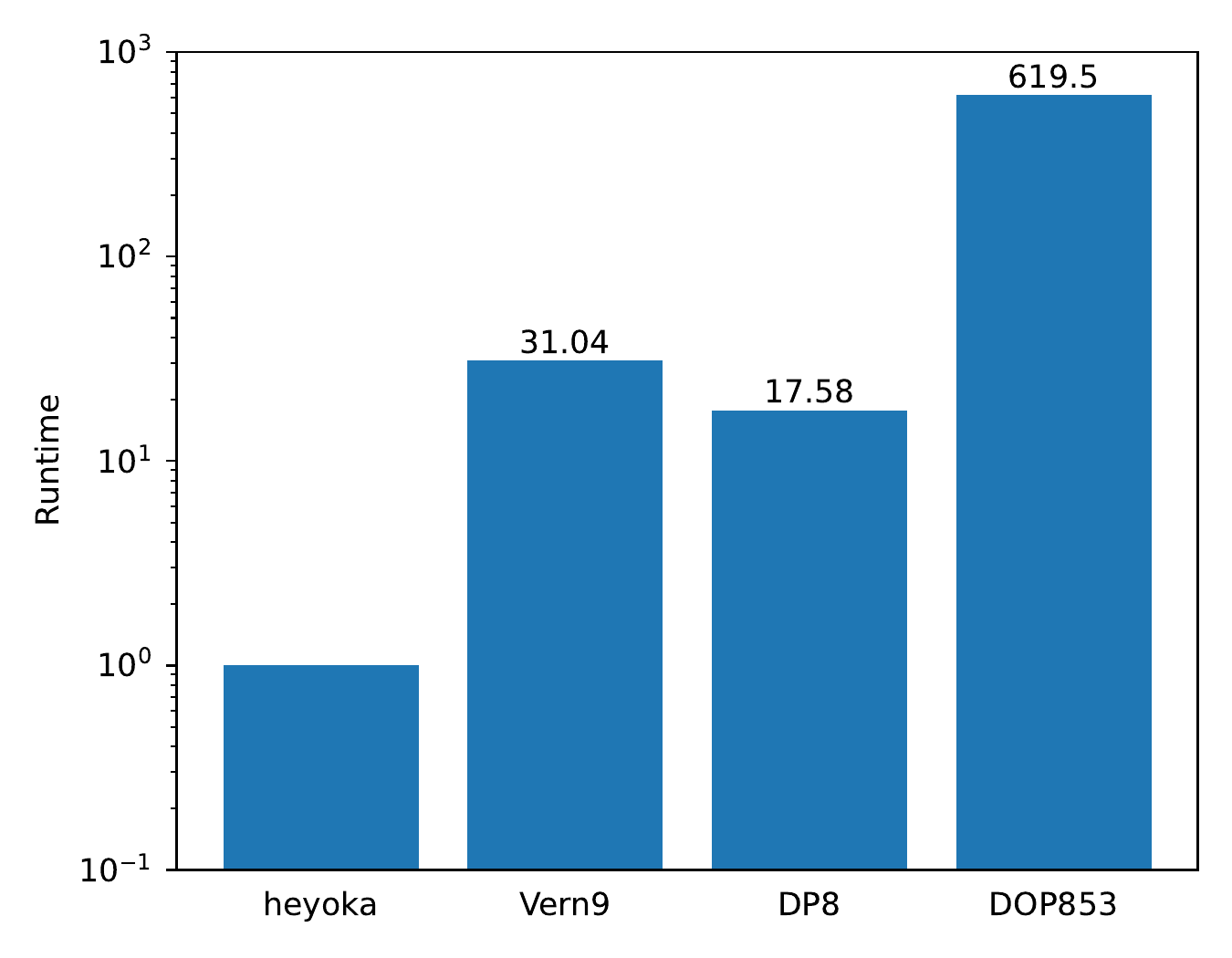}
 \caption{Runtime performance for the computation of a Poincar\'e section in the Hénon-Heiles dynamical system described in \S\ref{subsec:third_int}. 
 The vertical axis shows the time required by several numerical integration libraries to compute
 the same Poincar\'e section, normalised by the timing for our Taylor integrator \heyoka{}.
 \heyoka{} is compared to the explicit Runge-Kutta schemes \texttt{Vern9} and \texttt{DP8} from
 \texttt{DifferentialEquations.jl} \citep{diffeqjl} and \texttt{DOP853} from \texttt{SciPy} \citep{scipy}, which
 support event detection via the sign change method. The error tolerance is set to $10^{-15}$ for all integrators.
 The timings were taken on an AMD Ryzen 3950X CPU.}
 \label{fig:psec_perf}
\end{figure}

From the performance point of view, we report in Figure \ref{fig:psec_perf}
a runtime comparison between \heyoka{}, the \texttt{Vern9} and \texttt{DP8} integrators from
\texttt{DifferentialEquations.jl} \citep{diffeqjl}, and the \texttt{DOP853} integrator from
\texttt{SciPy} \citep{scipy}. The \texttt{DP8} and \texttt{DOP853} integrators are implementations
of the explicit Dormand-Prince Runge-Kutta 8/5/3 method \citep{hairer1993solving},
while \texttt{Vern9} is a ninth-order explicit Runge-Kutta scheme \citep{verner2014explicit}.
Apart from \heyoka{}, all integrators employ the sign change method for event detection.

In this simple test, we can see how \heyoka{} performs considerably better than the other integrators,
being more than an order of magnitude faster than the integrators from \texttt{DifferentialEquations.jl} and more than two
orders of magnitude faster than the \texttt{SciPy} integrator.
Regarding the \texttt{SciPy} integrator,
we should emphasise that its performance in this test is penalised by the slowness of the Python interpreter.
Another performance comparison with \texttt{DOP853} will be presented in \S\ref{subsec:ecl_traj},
where \texttt{DOP853} performs significantly better thanks to the fact that
the evaluation of the right-hand side of the ODEs is mostly done in compiled C code (rather than
in interpreted Python code).

\subsection{Collision detection in the outer Solar System}
\label{subsec:outer_ss_coll}
In this test, we will be considering the long-term dynamics of the outer Solar System while keeping track of possible collisions between the planets.
The dynamical system is set up as a gravitational 6-body problem consisting of the Sun, Jupiter, Saturn, Uranus, Neptune and Pluto, with initial conditions
taken from \citet{applegate1986outer}. This dynamical system does not experience close encounters (let alone collisions) 
between the planets over a timespan of (at least) $10^6\,y$.

Collision detection is implemented with 15 event equations of the form
\begin{equation}
\left(x_i-x_j\right)^2 + \left(y_i-y_j\right)^2 + \left(z_i-z_j\right)^2 - \left(2R\right)^2 = 0,\label{eq:ss_ev_eq00}
\end{equation}
where $\left( x_i, y_i, z_i \right)$ and $\left( x_j, y_j, z_j \right)$ are the Cartesian coordinates of the $i$-th and $j$-th body respectively,
and $R$ is the
planetary radius, which, for simplicity, is assumed to be Jupiter's radius for all bodies. The event equation \eqref{eq:ss_ev_eq00} is
satisfied when two planets (represented as spheres of radius $R$) come into contact.
The error tolerance $\varepsilon$ in this numerical experiment is set to $10^{-18}$: as shown in \citet{hey_mnras},
an error tolerance below machine epsilon ($\sim 10^{-16}$) allows our Taylor integrator to
satisfy Brouwer's law for the conservation of prime integrals over billions of dynamical timescales.

\begin{figure}
 \includegraphics[width=\columnwidth]{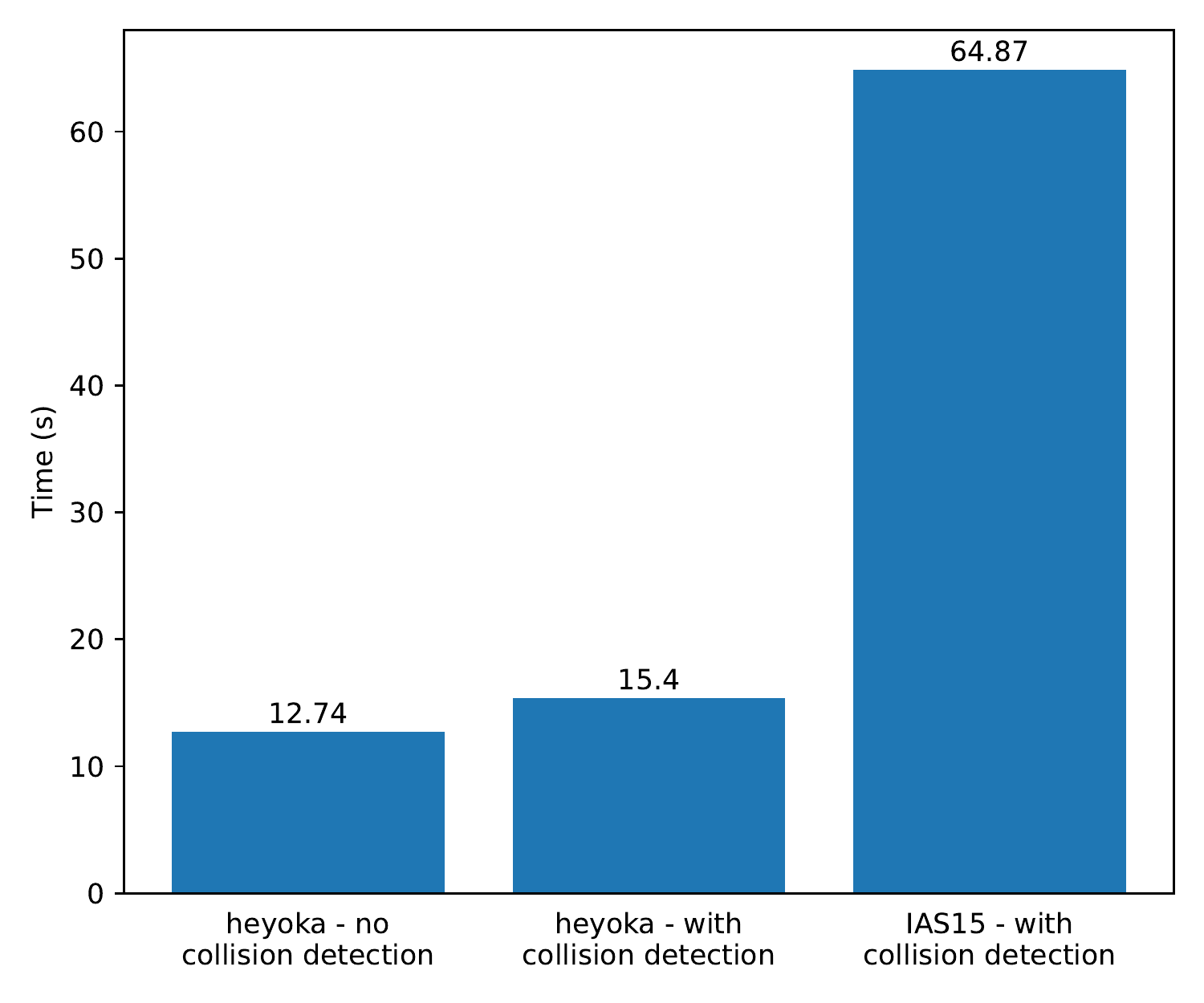}
 \caption{CPU timings for a $10^6\,\textnormal{y}$ integration of the outer Solar System (see \S\ref{subsec:outer_ss_coll}). Our Taylor integrator,
 \heyoka{}, is benchmarked with and without collision detection. The timing for the \texttt{IAS15} integrator from \texttt{REBOUND} are also included.
 Collision detection in \heyoka{} results in a $\sim 20\%$ slowdown with respect to an integration without collision detection. The timings were
 taken on an Intel Xeon Platinum 8360Y.}
 \label{fig:coll_perf}
\end{figure}

The purpose of this test is dual:
\begin{itemize}
    \item first, we want to measure the overhead of event detection in a situation where no event ever triggers;
    \item second, we want to verify the claims made \S\ref{subsec:complexity} and \S\ref{subsec:fast_excl} about the optimality of our root isolation approach.
    Specifically, we want to verify experimentally that the fast exclusion check described in \S\ref{subsec:fast_excl} is effective at quickly excluding the occurrence
    of events.
\end{itemize}
The CPU timings for a total integration time of $10^6\,\textnormal{y}$ are visualised in Figure \ref{fig:coll_perf}, where we also included the timing for
\texttt{IAS15} \citep{rein2015ias15}. Like \heyoka{}, \texttt{IAS15} is a non-symplectic N-body integrator which is nevertheless
able to satisfy Brouwer's law over
billions of dynamical timescales while at the same time resolving accurately the dynamics of close encounters.
\texttt{IAS15} supports collision detection via the usual approach of checking for sphere-sphere overlaps at the beginning of
each timestep. Although this method is very cheap computationally, it cannot guarantee that collisions are not missed.

The timings for \heyoka{} show how the overall performance overhead of event detection is $ \sim 20\%$ for this test. Regarding the behaviour of the root-finding
algorithm, in this test the fast exclusion
check described in \S\ref{subsec:fast_excl} was always able to exclude the triggering of collision events with no false positives.

\subsection{Collisions and close encounters between planetary embryos}
\label{subsec:embryos}
Whereas in the previous test (\S\ref{subsec:outer_ss_coll})
we examined a non-collisional N-body system, in this example we will test the behaviour of our event detection approach
in an N-body system characterised by frequent collisions and close encounters.

Collisional planetary systems play an important role in the study of the origin and evolution of the Solar System.
Current theories of planet formation posit that planetary embryos form via the collision of kilometer-sized planetesimals,
which in turn originate from the aggregation of dust in protoplanetary disks \citep{klahr2020turbulence}.
Planetary embryos have masses of about $10^{22}$ to $10^{23}\,\textnormal{kg}$,
and they have diameters up to a few thousand kilometers. Over time, planetary embryos collide and merge with each other, eventually resulting in the
formation of planets.

Numerical studies of planet formation (e.g., in the context of population synthesis, see \citet{mordasini2009extrasolar})
often employ hybrid integrators such as \texttt{MERCURY} \citep{1997DPS....29.2706C} or the closely-related
\texttt{MERCURIUS} \citep{rein2012rebound}. Hybrid integrators consist of a symplectic integration scheme paired to a non-symplectic, high-accuracy integrator.
The symplectic integrator is employed as long as the planets are not experiencing close encounters. When a close encounter is detected, the hybrid integrator
switches to the high-accuracy scheme, resolves the encounter and finally resumes the simulation with the symplectic integrator. Note that
although it is certainly possible to pair a Taylor integrator to a symplectic integrator to obtain a new type of hybrid integrator, in this example
we will use \heyoka{} to integrate both long-term dynamics and close encounters (i.e., without switches to other integration schemes).

Inspired by a numerical experiment outlined in \citet{chambers1999hybrid}, we set up a gravitational N-body system consisting of the Sun and 30 planetary
embryos with masses ranging between $0.2$ Earth masses and $0.6$ Lunar masses. The planetary embryos are placed on randomly-generated low-eccentricity and
low-inclination orbits between $0.5\,\textnormal{AU}$ and $1.2\,\textnormal{AU}$.
We follow the evolution of this dynamical system for $10^4$ years, and we keep track of both collisions and close encounters between the embryos.
Collisions are detected via event equations like \eqref{eq:ss_ev_eq00}. For the detection of close encounters we use the event equation
\begin{multline}
\frac{1}{2}\frac{d\left( \boldsymbol{x}_i - \boldsymbol{x}_j\right)^2}{dt}=
\left( x_i - x_j \right)\left( v_{xi} - v_{xj} \right) 
 \\+\left( y_i - y_j \right)\left( v_{yi} - v_{yj} \right)
 +\left( z_i - z_j \right)\left( v_{zi} - v_{zj} \right) = 0,
\end{multline}
which will be satisfied when the mutual distance between embryos $i$ and $j$ is at a stationary point. In order to filter out distance maxima, we impose the
additional requirement that the direction of the event (i.e., the time derivative of the left-hand side of the event equation) must be \emph{positive},
so that only distance minima are detected.

\begin{figure}
 \includegraphics[width=\columnwidth]{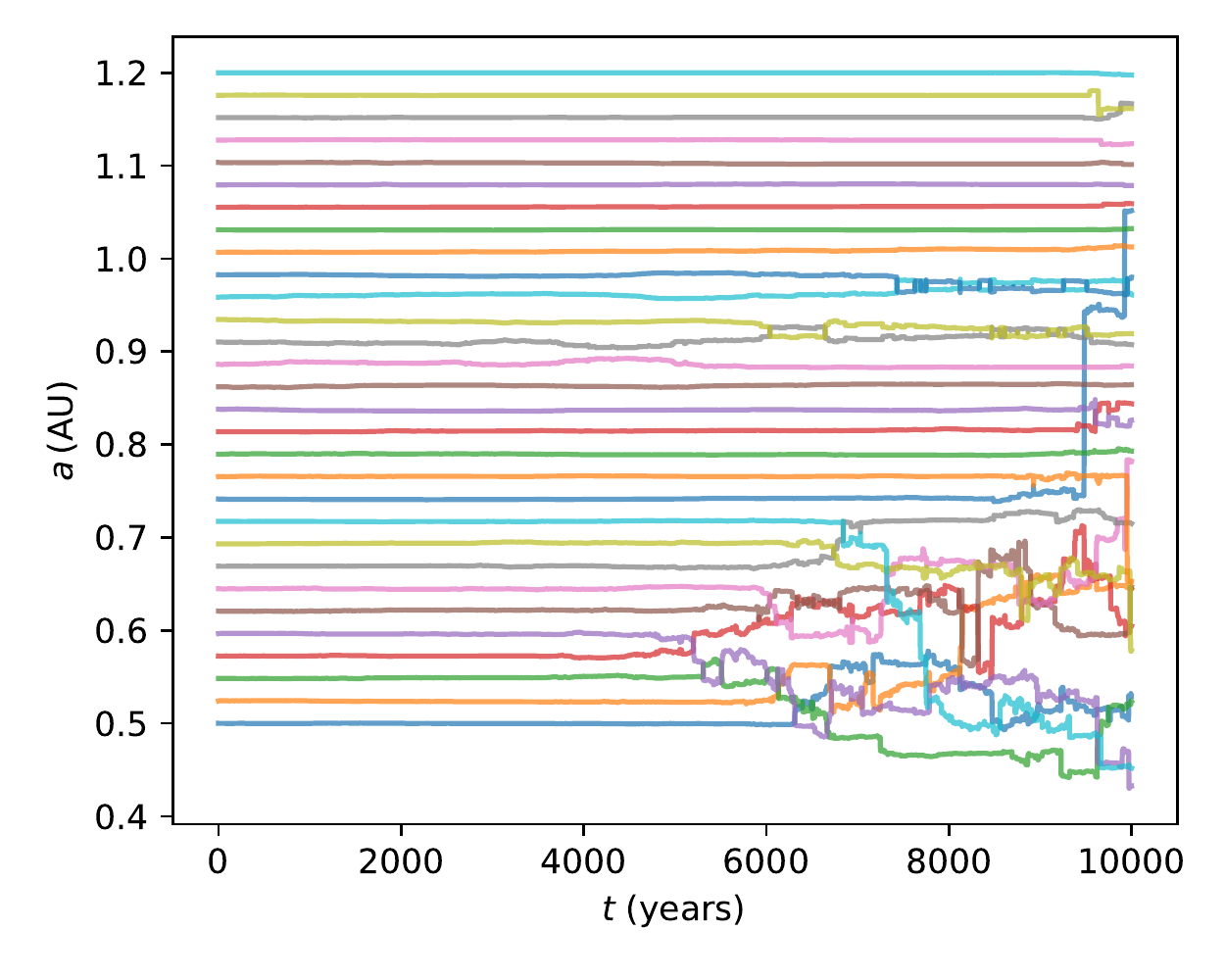}
 \caption{Evolution in time of the semi-major axes of 30 planetary embryos orbiting the Sun (see \S\ref{subsec:embryos}). The embryos are initially
 placed on quasi-circular quasi-planar orbits between $0.5\,\textnormal{AU}$ and $1.2\,\textnormal{AU}$. The chaotic gravitational interactions
 between the embryos eventually lead to close encounters and collisions.}
 \label{fig:embryos}
\end{figure}

The chaotic nature of this system is evident in the time evolution of the semi-major axes of the embryos (Figure \ref{fig:embryos}). Our event detection
algorithm detected a total of 31 planetary collisions within the $10^4$ years timespan.
The cumulative number of close encounters between the embryos as a function
of the encounter distance is visualised in Figure \ref{fig:embryos}, with the closest planetary encounter taking place at a mutual
distance of $\sim 2.8\,\textnormal{km}$. These results are in good agreement with the results from \citet{chambers1999hybrid}.

\begin{figure}
 \includegraphics[width=\columnwidth]{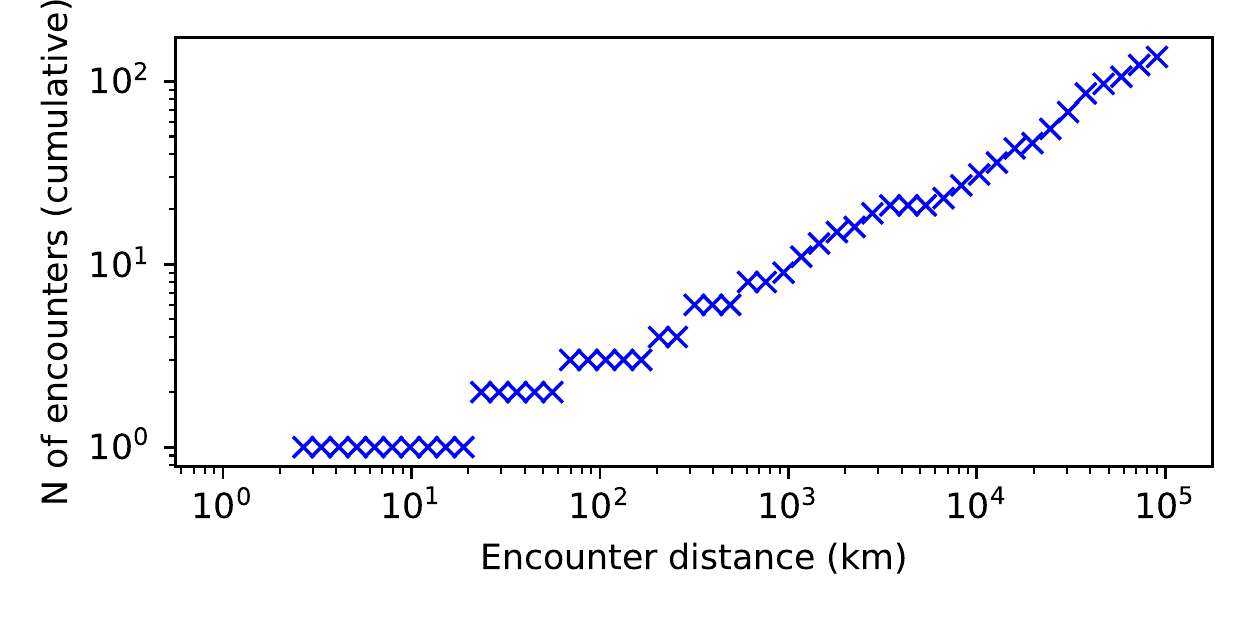}
 \caption{Cumulative number of close encounters between planetary embryos in the inner Solar System as a function of the encounter
 distance in the setup described in \S\ref{subsec:embryos}. The total simulation time is $10^4$ years.}
 \label{fig:c_enc}
\end{figure}

Regarding the performance of the polynomial root finding algorithm in this test, like in the previous example (\S \ref{subsec:outer_ss_coll})
the fast exclusion check
described in \S\ref{subsec:fast_excl} never resulted in false positives. When events did trigger, the polynomial root
isolation procedure always terminated in a single iteration. That is, the CA algorithm
immediately detected a single sign change in the coefficients of the transformed polynomial, thereby determining that
a single event would trigger within the timestep.
Thus, like in \S \ref{subsec:outer_ss_coll}, we conclude that our experiments indicate that the polynomial root finding
algorithm is behaving in the optimal way also in collisional planetary N-body simulations.

We need to emphasise that our goal in presenting this test is not necessarily
to advocate for the use of \heyoka{} in all simulations of this kind. When the collisional dynamics matters only statistically, the risk of missing a single collision is acceptable and existing hybrid integrators may be able to simulate the system more efficiently. 
We do believe though, also in those cases, that the event detection guarantees provided by \heyoka{} (coupled to its abilities to respect Brouwer's law in long-term integrations and to accurately resolve close encounters via its adaptive timestepping scheme) are useful to measure the magnitude of the effects introduced by allowing planetary collisions to be missed.

\subsection{Eclipsed trajectories around irregular bodies}
\label{subsec:ecl_traj}
In this last example, we consider the motion of a test mass around an irregular body,
subject to the gravity and to the solar radiation acceleration, accounting for possible eclipses.
Such a problem is relevant for the study of the long term stability of spacecraft orbits around asteroids and comets (as most recently argued by \cite{lang2021spacecraft}), as well as for the accurate prediction of possible ejecta from active asteroids such as the documented case of the asteroid Bennu \citep{chesley2020trajectory}.

Let us model the asteroid gravity using a mascon model. 
Mascon models are known to be appropriate to describe the gravity of small irregular bodies where they have the advantage of being computationally efficient \citep{wittick2017mascon} and able to represent also non homogeneous bodies \citep{park2010estimating}. 
We indicate with $N$ the number of mascons, with $m_j$ their masses and with $\mathbf r_j$ their position. 
Our reference frame is attached to the body, assumed to be in a uniform rotation with an angular velocity $\boldsymbol \omega$. Hence Coriolis and centrifugal accelerations need to be accounted for. 
We model the solar radiation pressure introducing an acceleration $\eta$ acting against the Sun direction $\hat{\mathbf i}_S$ and regulated by the eclipse factor $\nu(\mathbf r) \in [0,1]$ which determines the eclipse-light-penumbra regime.
In particular, $\nu=1$ when the body is fully illuminated, $\nu=0$ when fully eclipsed (umbra). 
Formally, the following set of differential equations are considered:
\begin{equation}
    \label{eq:eom_mascon}
\ddot {\mathbf r} = -G \sum_{j=0}^N \frac {m_j}{|\mathbf r - \mathbf r_j|^3} (\mathbf r - \mathbf r_j) - 2 \boldsymbol\omega \times \mathbf v - \boldsymbol \omega \times\boldsymbol\omega \times \mathbf r - \eta \nu(\mathbf r) \hat{\mathbf i}_S(t),
\end{equation}
where the Sun direction $\hat{\mathbf i}_S(t) = \mathbf R(t)\hat{\mathbf i}_S(0)$ is modelled, neglecting the asteroid orbital motion, rotating the initial Sun direction $\hat{\mathbf i}_S(0)$ around the asteroid rotation axis, of an angle $\omega t$. 
The eclipse factor $\nu(\mathbf r)$ in the above equation adds quite some complexity to the dynamics. 

Let us neglect here the penumbra regime and assume $\nu$ switches instantaneously between extreme values whenever the center of mass enters into the shadowed region created by a point source Sun.
The resulting dynamical system is then a hybrid system switching dynamical regimes according to whether the spacecraft is subject or not to the solar radiation acceleration. 
In order to detect such a switch condition and be able to use the event detection machinery of any initial value solver, we introduce a function changing sign when entering/exiting eclipsed areas. The eclipse function $F(\mathbf r, \hat{\mathbf i}_S)$ is defined as the distance of $\mathbf r$ from the shadow cone (degenerated into a cylinder in this case) cast by the body along the $\hat{\mathbf i}_S$ direction, if $\mathbf r$ is outside the cone. 
Whenever $\mathbf r$ is inside the eclipsed portion of the space around the irregular body, the eclipse function is defined as the length of the segments belonging to the ray passing through $\mathbf r$ that are inside the body, transformed into a negative number. We have built a function whose zero level curve, fixing any direction, describes the border of the eclipsed area and can be thus detected as an event and used to propagate the hybrid system.
\begin{figure}
 \includegraphics[width=\columnwidth]{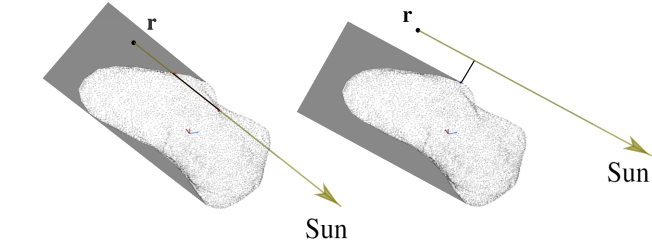}
 \caption{Definition of the eclipse function $F(\mathbf r, \hat{\mathbf i}_S)$ (black segment) in the case of the asteroid Eros. The eclipse function is defined as a positive number outside of the eclipse cone (right). Inside the eclipse cone (left) is a negative number whose magnitude is the length of the ray inside the body.}
 \label{fig:ef}
\end{figure}
In Figure \ref{fig:ef}, a visualization of the eclipse function $F$ is reported. 

Assuming a shape model is available for the irregular body, its three dimensional surface triangular mesh can be used to compute the numerical value of the eclipse function by applying, for example, the Möller-Trumbore algorithm \citep{moller1997fast} to the ray originating in $\mathbf r$ and directed as $\hat{\mathbf i}_S$. An alternative approach to compute the eclipse function would be to use ray marching \cite{sherstyuk1999fast} and rely on a distance function computed from the available surface triangular mesh.
These approaches to computing the eclipse function have two main drawbacks.
Firstly, they can be computational expensive. 
Depending on the accuracy of the shape model, the number of triangles in the mesh is typically high, and even for a relatively low precision model, anyway in the order of thousands. 
The computation of the eclipse function is to be done at each time step of the numerical integration and therefore adds up directly to the computational complexity of a single integration step. 
Secondly, nor the Möller-Trumbore algorithm nor the ray marching approach are differentiable to an arbitrary order: if we are to take advantage of the properties of the proposed event detection method in a Taylor integration scheme, the event function needs to have this property.
For this purpose we represent the eclipse function with a feedforward artificial neural network relying on its universal function approximator property \citep{calin2020universal}. 
The purpose of this test is thus threefold:
\begin{itemize}
    \item first, we want show the use of \heyoka\ in cases where the event equations and/or the dynamical equations contain a deep network. The highly recursive nature of deep networks allows the just-in-time compilation to happen efficiently even if the explicit mathematical expressions involved are not manageable;
    \item second, we want to show a practical case where the use of a reliable event detection machinery makes a significant difference on the precision of the obtained results;
    \item third, we want to verify the claims made \S\ref{subsec:d_stick} on the discontinuity sticking problem.
\end{itemize}

The use of deep networks to construct three dimensional differentiable representations of complex objects has been demonstrated by a number of recent works \citep{mildenhall2020nerf, park2019deepsdf, izzo2021geodesy, derksen2021shadow} and is often referred to as implicit neural representation. The eclipse function we here introduced, is similar to the signed distance function used by \cite{park2019deepsdf} to learn implicit neural representations of objects. Here we do not discuss in detail the generic network architecture employed, its training and performances which are, instead, the subject of a future separated work.
For the purpose of this paper we use one of such networks $\mathcal N$, that we call eclipseNet, trained to approximate the eclipse function over the shape model of the asteroid Eros taken from the work of \cite{erospoly} and reduced to 14744 triangles for convenience.
At the end of the training the average mean square error obtained for this particular eclipseNet translates to less than 1\% relative error on the prediction of the zeros of the eclipse function, and a three orders of magnitude speed up with respect to the approach based on the Möller-Trumbore algorithm. The resulting eclipse function is fully differentiable (the network non-linearities used are hyperbolic tangents) at arbitrary order, so that it is possible to compute, along any trajectory solution to eq. \eqref{eq:eom_mascon}, the quantities
$$
\frac{d^n F}{dt^n} = \frac{d^n \mathcal N(\mathbf r(t), \hat{\mathbf i}_S(t))}{d t^n},
$$
exploiting the automated differentiation machinery of the Taylor integrator.
\begin{figure}
 \includegraphics[width=\columnwidth]{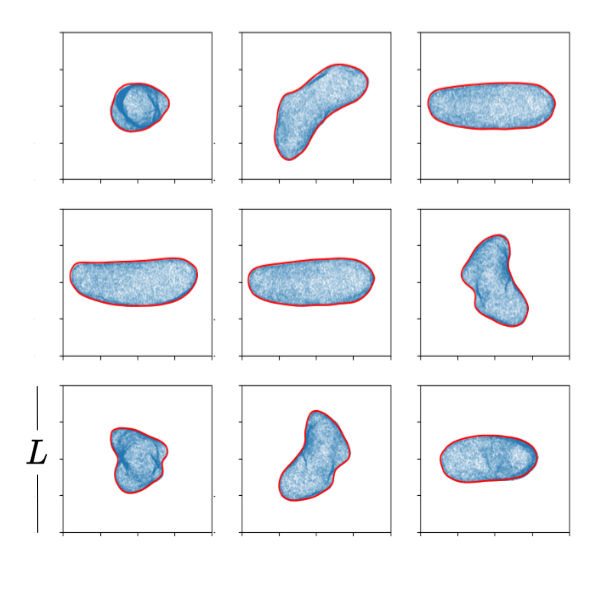}
  \caption{Visualization of the eclipseNet inference over 9 different Sun directions unseen during training. The red line represents the zero level curve of the eclipseNet which is to be detected by the event machinery signalling the entrance (or exit) into the asteroid shadow cone. In blue we report the vertices of the shape model for Eros projected onto the view plane. Length units are $L = 19.52$ km}
 \label{fig:eclipsenet_validation}
\end{figure}
To get a visual glimpse on the learned differentiable version of the eclipse function, we visualize, in Figure \ref{fig:eclipsenet_validation}, the network predictions from nine random views $\hat{\mathbf i}_S$ not encountered during the training. In other words, the network was never exposed to any of these data during training and yet is able to reproduce consistently to good detail the eclipse function next to its zero.

We now move on and consider again the initial value problem for eq. \eqref{eq:eom_mascon} where the eclipse factor $\nu$ switches between its two extreme values in correspondence to the roots of the event equation defined by the eclipseNet: $\mathcal N(\mathbf r, \hat{\mathbf i}_S) = 0$. The sign of the event equation derivative will determine whether the spacecraft is entering (negative) or leaving (positive) the eclipse.
\begin{figure}
 \includegraphics[width=\columnwidth]{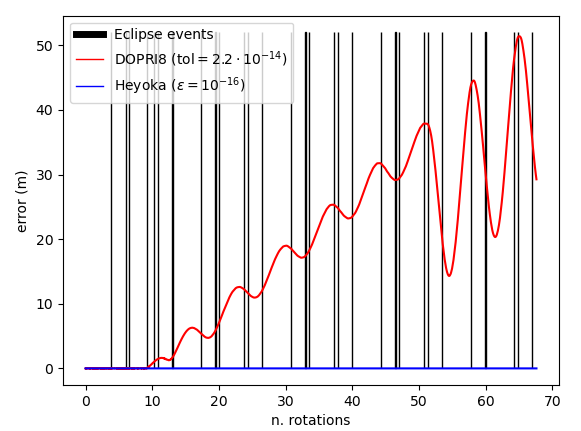}
  \caption{As Eros rotates and the orbit is propagated, a number of eclipse events are encountered. The effect of missing some of the events is here visualized in terms of the resulting error. As some early eclipses are missed by the event detection based on a 8th order Dormand-Prince, the resulting error grows invalidating the outcome. On the other hand, \heyoka{}'s event detection keeps the usual Taylor integration accuracy through the various eclipse events.}
 \label{fig:eclipse_error}
\end{figure}
We solve the same initial value problems using our Taylor integrator \heyoka{} (where the switching of $\nu$ between the extremal values is implemented in the callback of a terminal event, see \S\ref{subsec:t_nt_ev}), and a standard Dormand-Prince 8th order numerical integrator (\texttt{DOP853}) from \texttt{SciPy} \citep{scipy} using a sign based event detection.
Like most modern packages for event detection, \texttt{DOP853} only monitors the sign of the event function along the numerical propagation searching for a change in its value as computed at two subsequent timesteps.
Upon triggering of such a condition a root finding method is deployed in that interval to find and return the zero of the event function. 
It is important to note here that solvers choose their adaptive step size so as to compute the dynamics accurately, without accounting for the event function variations \citep{shampine2000event}. 
As a consequence, there is no way to know if events are lost along the propagation and one is often left to hope that setting the accuracy to the maximum possible value the likelihood of significant event being lost is small if not vanishing. 

In our tests, we set the accuracy of the Dormand-Prince numerical integrator, both in terms of absolute and relative tolerances, to the highest level allowed (100 times the machine epsilon), so that the chances to lose events are minimized. 
We perform simulations first without tracking events and then activating the event machinery and thus simulating also the eclipsed dynamics. We propagate the same initial conditions in different setups and use non dimensional units setting the Cavendish constant to one, using $L=19.52$ km as length unit and $M=6.687\cdot 10^{15}$ kg as mass unit. The initial conditions simulated are $\mathbf r_0 = [0, 3, 0], \mathbf v_0 = [-\sqrt 3 / 3  \cos i + 3\omega, 0, \sqrt3 / 3 \sin i]$, where $i$ is the initial inclination here set to $21^\circ$ and $\omega$ is set to . Different choices of the initial conditions return similar results as far as they allow for eclipses to be present. The integration time is set to $T = 100 \pi$. 

\begin{figure}
 \includegraphics[width=\columnwidth]{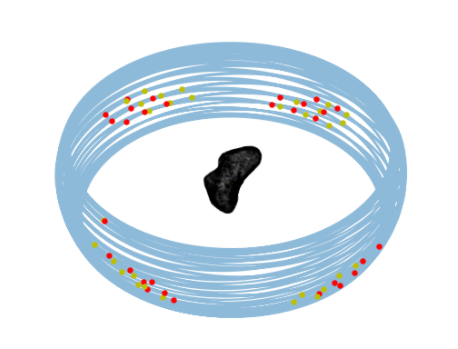}
  \caption{Simulated orbit around Eros. Points of entry into eclipse are indicated in red, while points of exit in yellow. Eclipses are encountered in different points around the orbit as the Sun moves in the asteroid reference frame.}
 \label{fig:eclipse_points}
\end{figure}

The results are reported in Table \ref{tab:doprivsheyoka}, where the different setups are compared in terms of CPU time and precision achieved. The maximum error reported refers to the position and is computed with respect to a ground truth generated by propagating the same system of differential equations using quadruple precision and the reliable event detection approach here introduced. In Figure \ref{fig:eclipse_points} we visualize the resulting orbit when eclipse events are reliably tracked.
\begin{table}
 \caption{Speed and precision achieved for several setups in the integration of eclipsed trajectories around irregular bodies (see \S\ref{subsec:ecl_traj}).
 When eclipse events need to be tracked \texttt{DOP853} is faster as it ignores the event equation in its adaptive stepper,
 thus taking larger steps at the price of missing out entire eclipse events.}
 \label{tab:doprivsheyoka}
 \begin{tabular}{lcc}
  \toprule
   & CPU time & Max. Error (position)\\
  \midrule
  \multicolumn{3}{c}{without eclipse events}\\
  \heyoka{} ($\epsilon=10^{-16}$)& 4.64s  & 2.44e-14\\
  \heyoka{} ($\epsilon=10^{-13}$)& 2.85s  & 1.49e-11\\
  \texttt{DOP853} (tol$=2.2\cdot10^{-14}$) & 8.5s & 5.26e-11\\
  \midrule
  \multicolumn{3}{c}{with eclipse events}\\
  \heyoka{} ($\epsilon=10^{-16}$)& 35.85s  & 1.23e-14\\
  \texttt{DOP853} (tol$=2.2\cdot10^{-14}$) & 6.57s & 1.32e-02\\
  \bottomrule
 \end{tabular}
\end{table}
The numerical propagations without eclipse events confirm, as reported already by \cite{hey_mnras}, that the Taylor integration scheme achieves great precision and CPU timings outperforming \texttt{DOP853}. When eclipse event need to be tracked \texttt{DOP853} becomes faster as it ignores the event equation in its adaptive stepper, thus taking larger steps at the price of missing out entire eclipse events. The consequences on the integration error are significant with ten orders of magnitude lost as eclipse events are missed. The reliable event machinery proposed in this paper and implemented in the Taylor integrator in \heyoka{}, is instead able to adaptively choose the step size as to follow also the event function with he necessary accuracy. This can slow down the integration, as more steps may be needed, but is a necessary and fair cost to pay to not introduce unacceptable errors (as explained in \S\ref{subsec:t_select}).

Regarding the approach proposed in \S\ref{subsec:d_stick} for the handling of discontinuity sticking
via automatic cooldown deduction in terminal events, our testing confirmed that all terminal events in \heyoka{} for this test
triggered exactly
once. Removing the automatic cooldown deduction heuristic also
confirmed that the eclipse crossing events often triggered multiple times in rapid
succession, due to the discontinuity sticking phenomenon (although they never led to an endless loop of re-detection of the same event).
By contrast, in this test the \texttt{DOP853} integrator would routinely get stuck in an infinite loop of re-detection of the eclipse
crossing events, leading to the need to implement a workaround involving the redefinition of the dynamical system at each eclipse
crossing and the manual specification of alternating directions for the event.

\section{Caveats and limitations}
\label{sec:caveats}
Despite the fact that our tests and numerical experiments indicate that our approach for the detection of events
in Taylor integrators via real-root isolation is both efficient and reliable, we need to point out a couple of caveats and limitations.

To begin with, our approach requires the event equations \eqref{eq:ev_eq00}
to be formulated as differentiable mathematical expressions of the state variables
(and, possibly, time). This is a limitation that stems directly from the very nature of Taylor integrators implemented on top of automatic differentiation,
which cannot deal with non-differentiable
expressions (such as, e.g., black box functions) in the definition of an ODE system. Such a limitation is though alleviated as the development of differentiable models approximating complex numerical algorithms (e.g., used to write the r.h.s. of the equations to be integrated) is often possible training, for example, artificial neural networks models as hinted in Section \S\ref{subsec:ecl_traj}.

Secondly, like most polynomial root-finding procedures, the real-root isolation process described in \S\ref{sec:rr_isol} breaks down
in correspondence of roots of multiplicity greater than one. As pointed out by \citet{park1996state}, it is in general exceedingly unlikely for
the high-order polynomials generated during a numerical integration to feature repeated roots (especially when using floating-point arithmetic).
Badly-conditioned event equations can, however, always lead to repeated roots in the Taylor series expansions. For instance, an event equation
of the type
\begin{equation}
\left[g\left( t, \boldsymbol{x} \left( t \right) \right)\right]^2 = 0
\end{equation}
will be troublesome, because both the event function and its time derivative will be zero when the event triggers.
This will translate to a Taylor series with a double root in correspondence of the event trigger time, which, at best, will result in reduced performance and, 
at worst, in missing events altogether. Additionally, in case of terminal events the automatically-deduced cooldown value in correspondence of a double root 
will tend to infinity (see \S\ref{subsec:d_stick}). Thus, as a general rule, event functions in which the event trigger times are stationary points are
to be avoided.

\section{Conclusions}
\label{sec:concl}
In this paper we introduced a novel, reliable approach for event detection in the numerical solution of ordinary differential equations via Taylor's method.
Our method is based on the construction of high-order polynomial approximations of the event functions, whose roots are located via
the Collins-Akritas real-root isolation scheme. Leveraging the availability of free dense output in Taylor's integration method,
our approach is computationally cheap while at the same time offering strong event detection guarantees.

We have tested the implementation of the new algorithm in our open source numerical integration package \heyoka{} on several astrodynamical problems of interest,
such as collisional N-body systems, spacecraft dynamics around irregular bodies accounting for eclipses and the computation of Poincar\'e{} sections.
In all cases the reliable event detection mechanism here proposed, coupled with the  known
properties of Taylor methods, produces results that are a significant improvement over the currently available alternatives.

Our work fills a gap in the capabilities of modern astrodynamical numerical integration packages, which typically do not feature
builtin event detection capabilities. We hope that our work will be useful to practitioners and researchers,
and that it will stimulate and enable further research and advances in computational celestial mechanics.

\section*{Acknowledgements}

This work is supported by the Deutsche Forschungsgemeinschaft (DFG, German Research Foundation) under Germany's
Excellence Strategy EXC 2181/1 - 390900948 (the Heidelberg STRUCTURES Excellence Cluster).

\section*{Data Availability}

No new data were generated or analysed in support of this research.

The source code of the software described in the paper
is freely available under an open source license at the repositories \url{https://github.com/bluescarni/heyoka} and
\url{https://github.com/bluescarni/heyoka.py}.

\bibliographystyle{mnras}
\bibliography{rr_tm}




\bsp	
\label{lastpage}
\end{document}